 \documentclass[pmlr,twocolumn,10pt]{jmlr} % W&CP article

\usepackage{booktabs}
\usepackage{siunitx}

\theorembodyfont{\upshape}
\theoremheaderfont{\scshape}
\theorempostheader{:}
\theoremsep{\newline}

\jmlrvolume{LEAVE UNSET}
\jmlryear{2023}
\jmlrsubmitted{LEAVE UNSET}
\jmlrpublished{LEAVE UNSET}
\jmlrworkshop{Machine Learning for Health (ML4H) 2023} % W&CP title

\title[Generating Drug Repurposing Hypotheses through the Combination of Disease-Specific Hypergraphs]{Generating Drug Repurposing Hypotheses through the Combination of Disease-Specific Hypergraphs}
\author{%
\Name{Ayush Jain} \Email{a.jain@duke.edu}\\
\addr Duke University, USA; Broad Instititue of MIT and Harvard, USA
\AND
\Name{Marie Laure-Charpignon} \Email{mcharpig@mit.edu}\\
\addr MIT, USA; Broad Institute of MIT and Harvard, USA
\AND
\AND
\Name{Irene Y. Chen} \Email{iychen@berkeley.edu}\\
\addr UC Berkeley, USA; UC San Francisco, USA
\AND
\Name{Anthony Philippakis} \Email{aphilipp@broadinstitute.org}\\
\addr Broad Institute of MIT and Harvard; USA
\AND
\Name{Ahmed Alaa} \Email{amalaa@berkeley.edu}\\
\addr UC Berkeley, USA; UC San Francisco, USA
}

\begin{document}

\maketitle

\begin{abstract}
The drug development pipeline for a new compound can last 10-20 years and cost over \$10 billion. Drug repurposing offers a more time- and cost-effective alternative. Computational approaches based on biomedical knowledge graph representations, comprising a mixture of nodes (e.g., disease, drug, protein, symptom, side effect) and their interactions, have recently yielded new drug repurposing hypotheses, including suitable candidates for COVID-19, such as remdesivir. In this study, we present a novel, disease-specific hypergraph representation learning technique to derive contextual embeddings of biological pathways of various lengths but that all start at any given drug and all end at the disease of interest. Further, we extend this method to multi-disease hypergraphs. Specifically, we combine the hypergraph of the disease of interest with those of its main associated risk factors to guide the search of repurposing candidates towards likely relevant drug classes. To determine the repurposing potential of each of the 1,522 drugs, we derive drug-specific distributions of cosine similarity values and ultimately consider the median for ranking. Cosine similarity values are computed between (1) all biological pathways starting at the considered drug and ending at the disease of interest and (2) all biological pathways starting at drugs currently prescribed against that disease and ending similarly at the disease of interest. We illustrate our approach with Alzheimer’s disease (AD), which affects over 55 million patients worldwide but still has no cure, and two of its risk factors: hypertension (HTN) and type 2 diabetes (T2D). We compare each drug’s rank across four hypergraph settings (single- or multi-disease): AD only, AD + HTN, AD + T2D, and AD + HTN + T2D. Notably, our framework led to the identification of two promising drugs, i.e., whose repurposing potential was significantly higher in hypergraphs combining two diseases: dapagliflozin (antidiabetic; moved up, from top 32\% to top 7\%, across all considered drugs) and debrisoquine (antihypertensive; moved up, from top 76\% to top 23\%). Our approach serves as a hypothesis generation tool, to be paired with a validation pipeline relying on laboratory experiments and semi-automated parsing of the biomedical literature.

%Two notable drugs that moved up significantly in ranks when their respective antagonist diseases were added were dapagliflozin (an antidiabetic, sodium-glucose cotransporter 2, moves from the top 32\% to the top 7\% across all considered drugs) and debrisoquine (an antihypertensive, sodium-dependent noradrenaline transporter, moves from the top 76\% to the top 23\%).
\end{abstract}

\section{Introduction}
\label{sec:introduction}
The development of new drugs can take more than 15 years, from the discovery and pre-clinical phase to review by regulatory agencies (\cite{xue_review_2018}). Hence, repurposing drugs already approved by the Food and Drug Administration or European Medicines Agency serves as a convenient alternative since they have already known to be safe in human populations. From a research and development perspective, drug repurposing is a less risky enterprise. Indeed, following compound identification, repositioned drugs would generally hit the market in less than 10 years. Beyond time savings, this strategy brings significant cost savings, potentially reducing the average pharmaceutical pipeline’s budget by over \$5 billion compared to traditional drug development. To date, drug repurposing encompasses three main approaches: computational biomedicine (\cite{jarada_review_2020}), biological experimentation, and their combination, e.g., through systems pharmacology (\cite{zhao_systems_2012}).

Computational approaches are both more time-effective and cost-effective than \textit{in vitro} or \textit{in vivo} biological experiments, which involve high-throughput screening or phenotypic screening based on animal and human models, respectively. Examples of available strategies include signature matching, genome-wide association studies, and the retrospective analysis of real-world clinical information  (\cite{wu_integrating_2022}). Their use has been unlocked by the concurrent emergence of technical advances such as biological microarrays and the increase in data accessibility, as illustrated by the rapid growth of electronic health records and biobanks  (\cite{tan_drug_2023}).

Simultaneously, massive genomic databases and cell lines have yielded 20+ high-quality biological and biomedical knowledge graphs (KG) such as the Hetionet (\cite{himmelstein_hetnet_2023}), SPOKE  (\cite{SPOKE_citation}), and PrimeKG  (\cite{chandak_building_2023}) and aggregating platforms such as the KG-Hub (\cite{noauthor_kg-hubbuilding_nodate}) to ensure that the former can be shared and made interoperable for downstream graph machine learning tasks. Network-based methods for drug repurposing rely on the encoding of interactions between entities (i.e., drugs, diseases, proteins, biological functions) that can be heterogeneous (i.e., inhibition, binding). These representations can help address both predictive (e.g., polypharmacy side effects) and inferential (e.g., reasoning over causal pathways) questions. Prior graph representations such as the multi-scale interactome (MSI)  (\cite{ruiz_identification_2021}) have proved useful in identifying known drug repurposing agents and formulating potential candidates. 
It has previously been shown how a disease-specific hypergraph representation learning technique could identify likely repurposing targets that were missed by the multiscale interactome (\cite{jain_hypergraph_2023}). Building on the promise of disease-specific hypergraph representation learning for the identification of suitable drug repurposing candidates by biological pathway similarity search, we propose to combine knowledge graph information pertaining to a disease of interest and comorbid diseases. The objective is to assess how this ``perturbation" may affect the ranks of drugs currently prescribed to mitigate the repercussions of comorbidities. Specifically, we hypothesize that combining disease-specific hypergraphs of comorbid diseases (e.g., HTN and T2D) with that of the condition of interest (e.g., AD) will boost the ranks of their respective antagonist drugs (antihypertensives and antidiabetics, respectively) upwards. Our findings will hopefully support the design of disease-specific network representation models and yield new drug repurposing insights. Additionally, our framework could enable precision medicine through combined hypergraphs of increasing granularity that closely match the disease profile of a pre-specified patient population. 

\section{Methods}
\label{sec:methods}
\subsection{Hypergraph Construction and Combination}
The Hetionet is accompanied by a graph database that allows us to query all biological pathways starting at one of the 1,522 drugs and ending at a disease of interest. The output of any given query consists in a stratified list of pathways, where strata correspond to metapaths. A given metapath stratum comprises pathways with the same length and composition (i.e., ordering of entities); it may include from 1 to 100,000+ biological pathways. For instance, one metapath stratum might be defined as ``drug-protein-protein-disease". This functionality of the Hetionet facilitates the selection of the most meaningful biological pathway, i.e., those whose metapaths contain one or more proteins. We have two ways of constructing the hypergraph for a specific disease. If the number of biological pathways starting with a drug and ending with the disease of interest is lower than 10,000, then we keep all pathways. If this number is greater than 10,000, then we select only the top 10\% of pathways per metapath stratum, based on the direct weighted path count. This metric is used to rank biological pathway belonging to the same metapath stratum. Their significance is determined by the number of connections that each entity on the pathway has in the original Hetionet graph (\cite{himmelstein_hetnet_2023}). This first step yields a disease-specific subgraph of the aggregate heterogeneous knowledge graph (See \figureref{fig:construction} (a), (b), (d)). 

Then, we unify the resulting biological pathways as hyperedges to create a disease-specific hypergraph (See \figureref{fig:construction} (c) and (e)). Finally, we turn the disease-specific hypergraph into a weighted graph, in which nodes represent biological pathways and are connected by an edge if they share one or more middle entities (e.g., a protein). Any entity other than that at the start or end of the pathway qualifies as such, since by design pathways included in the hypergraph all start with a drug and end with the disease of interest. We reasoned that this edge definition, based on the similarity of middle entities, would capture functional relationships among biological pathways belonging to different metapath strata (see \figureref{fig:construction} (f)). We defined the edge weight \textit{w} as the number of middle entities shared by the two pathway nodes. Further, we re-scaled each weight for it to be in the interval (0,1]. Hypergraphs combining multiple diseases into a single structure would be constructed similarly: all disease-specific hypergraphs would be overlaid before final transformation into a unique weighted graph.

\begin{figure}[htbp]
 % Caption and label go in the first argument and the figure contents
 % go in the second argument
\floatconts
  {fig:construction}
  {\caption{Creating and combining two disease-specific hypergraphs into a weighted graph}}
  {\includegraphics[width=1\linewidth]{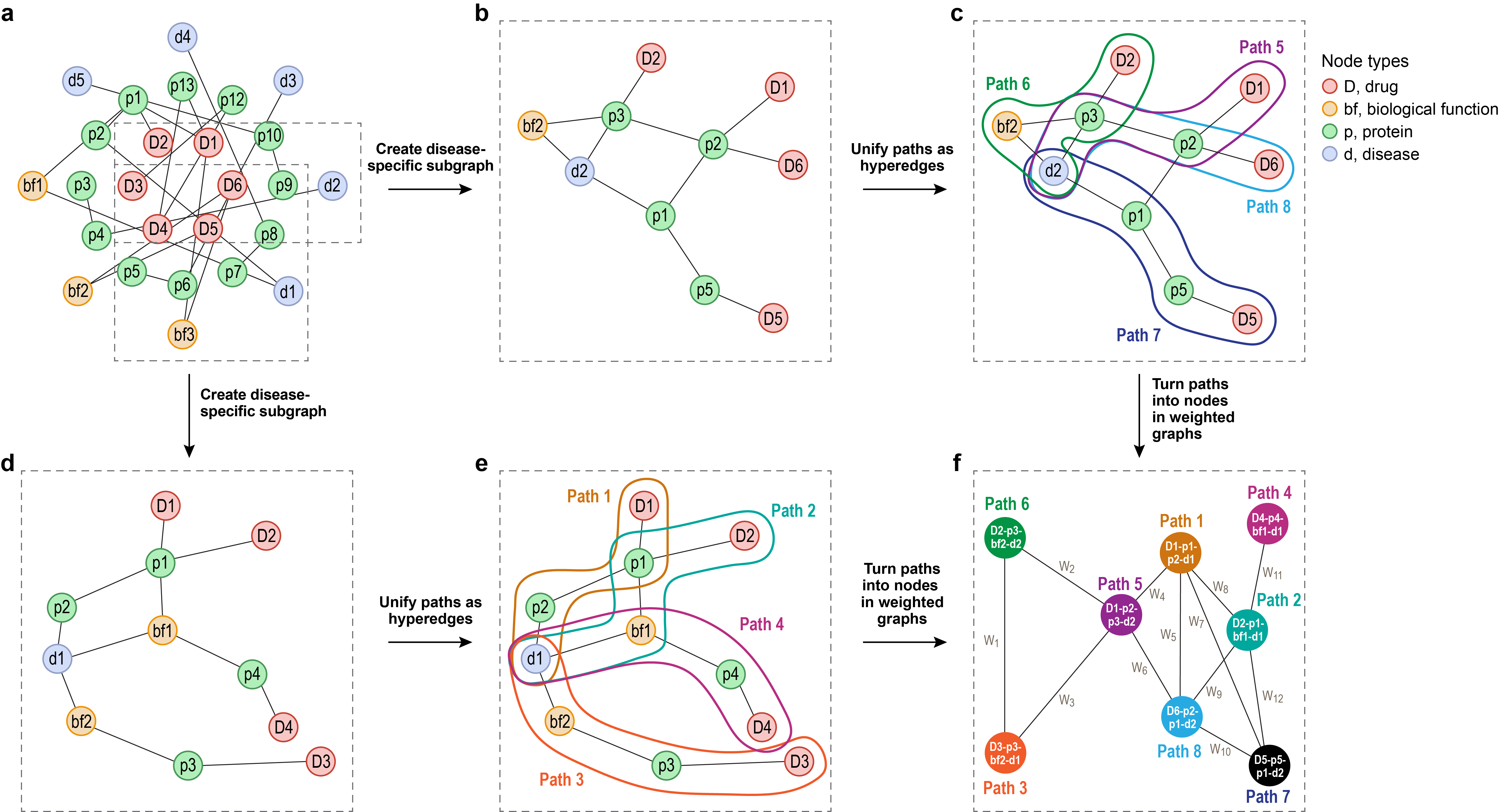}}
\end{figure}
\subsection{Hypergraph Representation Learning}
We learned biological pathway embeddings of size $n=64$ by initiating a random walk on the transformed graph shown in \figureref{fig:construction} (f). We accounted for the presence of weighted edges by sampling neighboring nodes proportionally to the strength of their connection. The random walker began at a selected node, then proceeded iteratively to an adjacent node chosen proportionally at random, and repeated this process for a predetermined number of steps. To account for uncertainty, each random walk in our transformed graph was replicated 10 times. For each replicate, the walk path length was set at 80 to ensure sufficient exploration of the weighted graph. In the discussion section, we outline additional sensitivity analyses to be performed to assess the robustness of our results to the choice of hyperparameters.

At each step of the random walk, the probability of transitioning from biological pathway $x$ to biological pathway $y$, given that the weighted edge between the two pathways is $w_{x,y}$ and that the total number of neighboring pathways connected to $x$ is $n_{x}$, is expressed as follows:

\begin{equation}\label{eq:walk}
P(v_i = y | v_{i-1} = x) = \frac{w_{x,y}}{\sum_{y=1}^{n_{x}}{w_{x,y}}}
\end{equation}

\subsubsection{Skip-Gram Model on Weighted Graphs}

By analogy with the dependencies present in natural language, we interpreted the resulting random walks as sentences. We utilized the Word2Vec Skip-Gram model provided by gensim to develop node embeddings for each biological pathway (\cite{grover_node2vec_2016}). This model predicts context words (nodes within the same walk) given a target word (a node). In the context of our disease-specific weighted graph, the embeddings of biological pathways learned through the random walk process encapsulate both local neighborhood structures and long-range connections among them. Subsequently, we use these embeddings for our pathway similarity search. We elaborate on the skip-gram part of our algorithm in \sectionref{apd:second}.

\subsection{Evaluating Path Embeddings}

Given a disease of interest, our study aimed to identify biological pathways analogous to those associated with drugs currently used to treat it or to mitigate its progression. In particular, we conducted a case study on Alzheimer's disease (AD) and considered medications prescribed to alleviate the symptoms and behavioral complications of AD. We focused primarily on three compounds: donepezil, galantamine, and memantine (reference drugs) (\cite{tan_efficacy_2014, howes_cardiovascular_2014, bond_effectiveness_2012}), approved by the FDA in 1996, 2001, and 2003, respectively. Additionally, we learned the embeddings of these same biological pathways on a larger hypergraph than the AD-only hypergraph, by combining the latter with those of two of the main risk factors for AD: hypertension (HTN) and type 2 diabetes (T2D). Our approach is disease-agnostic and can be extended to other diseases, upon the supply of a list of compounds currently used in clinical practice or previously suggested as repurposing candidates.

We represented the embedding of each biological pathway as a vector of size $n$: pathways starting at a drug were denoted as vector \(A\) and those ending at AD or a reference drug were denoted as vector \(B\).

We calculated the cosine similarity for each pair of \(A\) and \(B\) vectors, leading to a distribution of cosine similarity values for each drug. Subsequently, we derived the median cosine similarity for every drug and used this metric to rank each drug from 1 (highest median cosine similarity) to 1,552 (lowest median cosine similarity). 
Mathematically, our procedure articulates as follows. First, for any given drug \(d\), we compute its cosine similarity with each reference drug \(j\). This results in a set \(S_d\) of cosine similarity values for each drug \(d\):
\begin{equation}
S_d = \left\{ \text{sim}(A_d, B_j) \mid j \in \text{reference drugs} \right\}
\end{equation}
\begin{equation}
\text{sim}(A_d, B_j) = \frac{\sum_{i=1}^{n} A_{di} B_{ji}}{\sqrt{\sum_{i=1}^{n} A_{di}^2} \cdot \sqrt{\sum_{i=1}^{n} B_{ji}^2}}
\end{equation}

Second, we obtain the median value $m_d$ of set \(S_d\) for each drug \(d\) and use it to rank repurposing candidate drugs; the drug with the highest median for the disease of interest would be assigned a rank of 1. 

Third, drug rankings were compared across hypergraph settings (i.e., single- or multi-disease) to evaluate the effect of explicitly accounting for the comorbidities of a disease when learning biological pathway embeddings. Our datasets and code scripts are available in the supplementary materials. 

\section{Results}
\label{sec:results}
We built four distinct hypergraphs of increasing size: AD only, AD + HTN, AD + T2D, and AD + HTN + T2D. To measure the extent of differences in the resulting biological pathway embeddings, we quantified the percent of overlap among their top drug candidates for repurposing (ranked based on the median cosine similarity between (1) all pathways starting at a drug and ending at AD and (2) all pathways starting at donepezil, galantamine, or memantine and ending at AD), by increment of 1\%. Notably, differences appeared in the composition of the top 6\% when comparing in turn each of the three multi-disease hypergraphs to the AD-only one. Beyond this threshold, full overlap was reached in all combined hypergraphs (see \figureref{fig:nodes} (a)). 

In addition to graph embeddings, we compared the network properties of the four single- and multi-disease weighted graphs. While the clustering coefficient was high overall across all hypergraph settings (see \figureref{fig:nodes}(b)), its value was lower in the transformed weighted graphs combining a disease and its risk factors into a fused hypergraph than in the original, AD-only one. The fact that there was less triadic closure suggests that overlaying multiple disease-specific hypergraphs is effectively adding information and could help increase the relevance of top candidates for drug repurposing. Concurrently, the weighted graph density decreased as the number of diseases in the underlying hypergraph increased, from AD only to AD + HTN + T2D. The latter had the lowest density of all three combined weighted graphs (See \figureref{fig:nodes} (d)), a trend similar to that observed with the clustering coefficient and that could be explained by the addition of a large number of biological pathways specific to HTN and T2D. When contrasting graphs combining AD with only one of the risk factors, we found that the AD + T2D weighted graph had a lower density than its AD + HTN counterpart (\figureref{fig:nodes}(d)). This pattern differs from the ordering of clustering coefficients described above (\figureref{fig:nodes}(b)). Although T2D and HTN share common underlying pathways, a better understanding of the mechanistic relationships between them and with AD is still needed; each of these two risk factors affects the development of the other, with possible repercussions on AD onset (\cite{htn_t2d_connections}).

Lastly, we also compared the distributions of top gene nodes present on the biological pathways underlying single- and multi-disease hypergraphs (See \figureref{fig:nodes}(c)). We found limited variability across the four considered settings, besides the presence of SLC22A6 in the AD + T2D hypergraph and of COG2 in the AD + HTN + T2D one. Both are protein-coding genes involved in transport activity: sodium-dependent excretion of potentially toxic organic anions (\cite{GeneCardsSLC22A6}) and trafficking of Golgi enzymes, respectively (\cite{GeneCardsCOG2}). Interestingly, SLC22A6 is associated with the pathway of statins, which may delay AD onset (\cite{statins}). Furthermore, COG2 is associated with glycosylation disorder, corroborating the recent discovery of pathway-specific glycosylation alterations unique to AD (\cite{glycosylation}).

\begin{figure}[htbp]
 % Caption and label go in the first argument and the figure contents
 % go in the second argument
\floatconts
  {fig:nodes}
  {\caption{Comparison of Combined Hypergraphs}}
  {\includegraphics[width=1\linewidth]{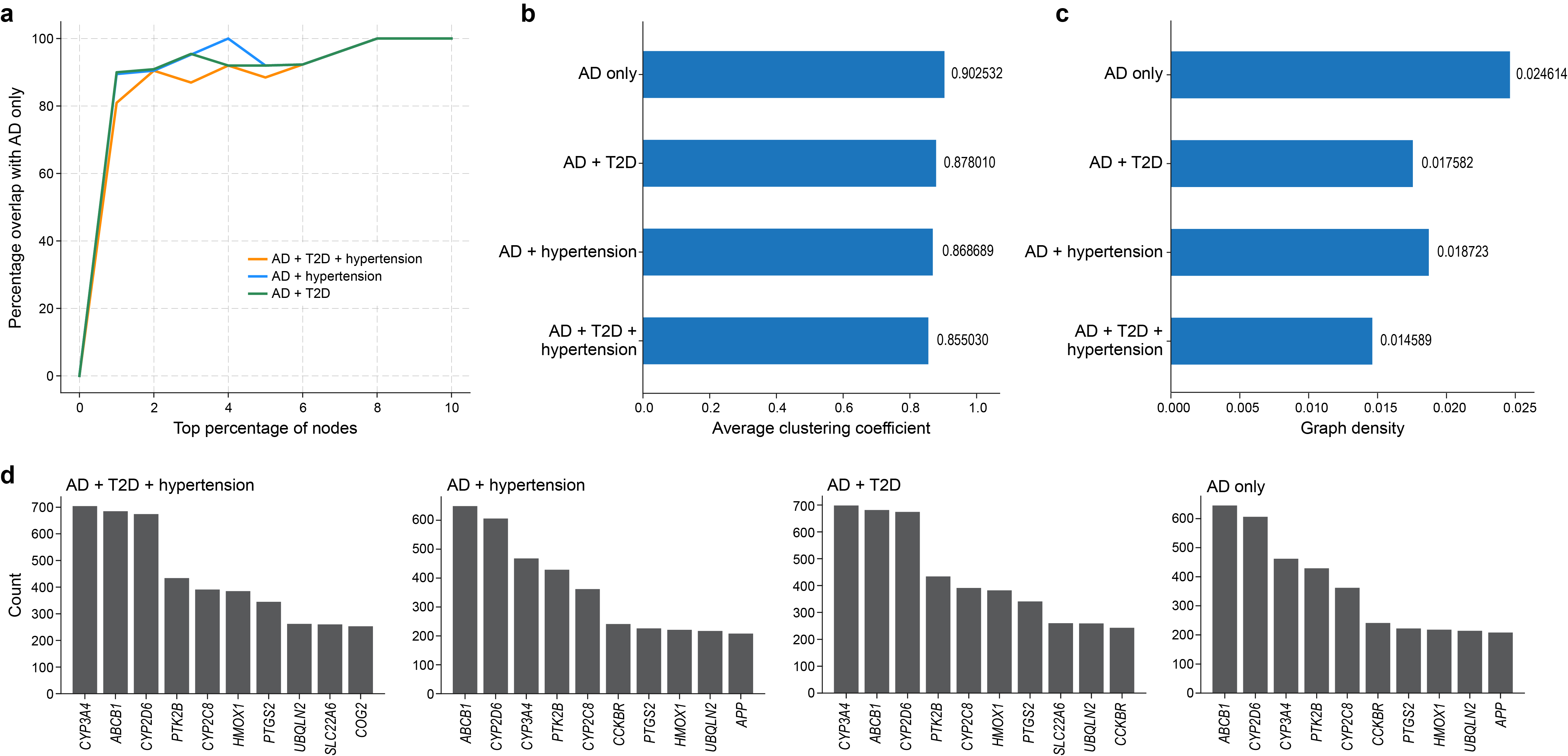}}
\end{figure}
%The largest median percentile increase among anti-hypertensives when the antagonist disease was added to the AD only hypergraph was 53\% by debrisoquine, a sodium-dependent noradrenaline transporter (moved from the top 76\% to the top 23\%). The largest median percentile increase among anti-diabetics was  25\% by dapagliflozin, a sodium-glucose cotransporter
%2 that moved from the top 32\% to the top 7\%. 
\section{Discussion \& Future Directions}
\label{sec:disco}

We proposed a novel biomedical hypergraph representation learning method to identify drug repurposing targets for a specific disease, through the combination of its hypergraph with those of one or several of its associated comorbidities. Taking the example of AD, we showed that our approach could help identify new drug repurposing candidates (see Appendix), upon analyzing top ranked compounds unique to the overlay of hypergraphs. Future work will consist in executing target trials using real-world data such as electronic health records and claims to test these hypotheses in patient populations with one or several risk factors. Additionally, we studied changes in the weighted graph properties (clustering and density) and gene composition of underlying biological pathways. The insights we gained about the involvement of the statin and glycosylation pathways, through the overlay of hypergraphs, corroborate some newly published conjectures, thus underscoring the hypothesis-generating role of our approach when deployed at scale. We plan to implement our method to the 800+ prevalent diseases referenced in the MSI and their associated comorbidities to formulate new therapeutic hypotheses--both single compounds and drug combinations; in doing so, we will provide pathway-based explanations to help elucidate their underlying mechanisms of action. 

Going forward, we will also complement our modified node2vec algorithm with other representation learning strategies to derive a set of embeddings for each biological pathway. Our current approach includes four hyperparameters: the fraction of top biological pathways included in the hypergraph when their number exceeds 10,000 (currently 10\%), the size of their embeddings (currently 64), the length of random walks on weighted graphs (currently 80), and the number of replicates (currently 10). We will run a large-scale sensitivity analysis to assess the robustness of our pathway similarity results to parameter changes and use Sobol indices to characterize the proportion of the variance in drug repurposing candidate ranks explained by each factor. 

%\acks{We would like to thank the Broad Summer Research Program and the Eric \& Wendy Schmidt Program PhD fellowship for funding passrts of this work.}

\bibliography{output}

\appendix

\section{Drug Rank Shifts Resulting From Hypergraph Combination}\label{apd:first}
When combining the AD-only hypergraph with those of two risk factors for this disease, HTN and T2D, we detected some significant changes in the ranks of certain antihypertensive and antidiabetic drugs. Notably, our framework led to the identification of two promising drugs, i.e., whose repurposing potential was significantly higher in two-disease combined hypergraphs than in the AD-only hypergraph: dapagliflozin (\cite{AlAdAwi2019}) (antidiabetic; moved up, from top 32\% to top 7\%, across all considered drugs) and debrisoquine (\cite{Brøsen1989}) (antihypertensive; moved up, from top 76\% to top 23\%). Interestingly, dapagliflozin and SGLT2 inhibitors more broadly have already shown repurposing potential for cardiovascular outcomes \citep{dapademonstratedpotential}. It may also contribute to a better control of risk factors such as excess body weight, elevated blood pressure, and dyslipidaemia, the prevention of nephropathy and retinopathy, and the mitigation of other diabetic complications (\cite{dapabenefits}). A trial started in February 2019 at Kansas University Medical Center aimed at determining whether dapagliflozin improved the brain's metabolism \citep{dapanewsarticle, dapatrialpage}. Furthermore, a recent cohort study in Ontario, Canada suggested that SGLT2 inhibitors were associated with lower dementia risk than active controls in older people with T2D \citep{diabetescare2023}. 

For a select set of drugs, \figureref{fig:ranks} shows the difference in ranks between the AD-only hypergraph and each of the three combined hypergraphs (AD + T2D, AD + HTN, or AD + T2D + HTN). For the two drug classes of interest (antidiabetics and antihypertensives), we report the drugs with the largest rank shifts, both negative (lower repurposing potential; in red) and positive (higher repurposing potential; in blue). Our approach could also be used to build drug repurposing portfolios specific to a given patient population, based on their set of comorbidities or known genetic variants putting them at higher risk for the disease of interest. 

In the future, we plan to emulate target trials among patient populations with T2D, HTN, or both to evaluate the comparative effectiveness of the drug repurposing candidates emanating from our hypergraph-based approach. If the drugs we identified are associated with a lower risk of AD than control drugs prescribed for the same indication, it would partially validate our hypothesis that the combination of comorbidity hypergraphs can accurately identify repurposing candidates in population with risk factors; actual randomized controlled trials would be needed to further confirm the benefits of our method.

\begin{figure}[htbp]
 % Caption and label go in the first argument and the figure contents
 % go in the second argument
\floatconts
  {fig:ranks}
  {\caption{Differences in drug ranks between the AD-only hypergraph and combined hypergraphs}}
  {\includegraphics[width=1\linewidth]{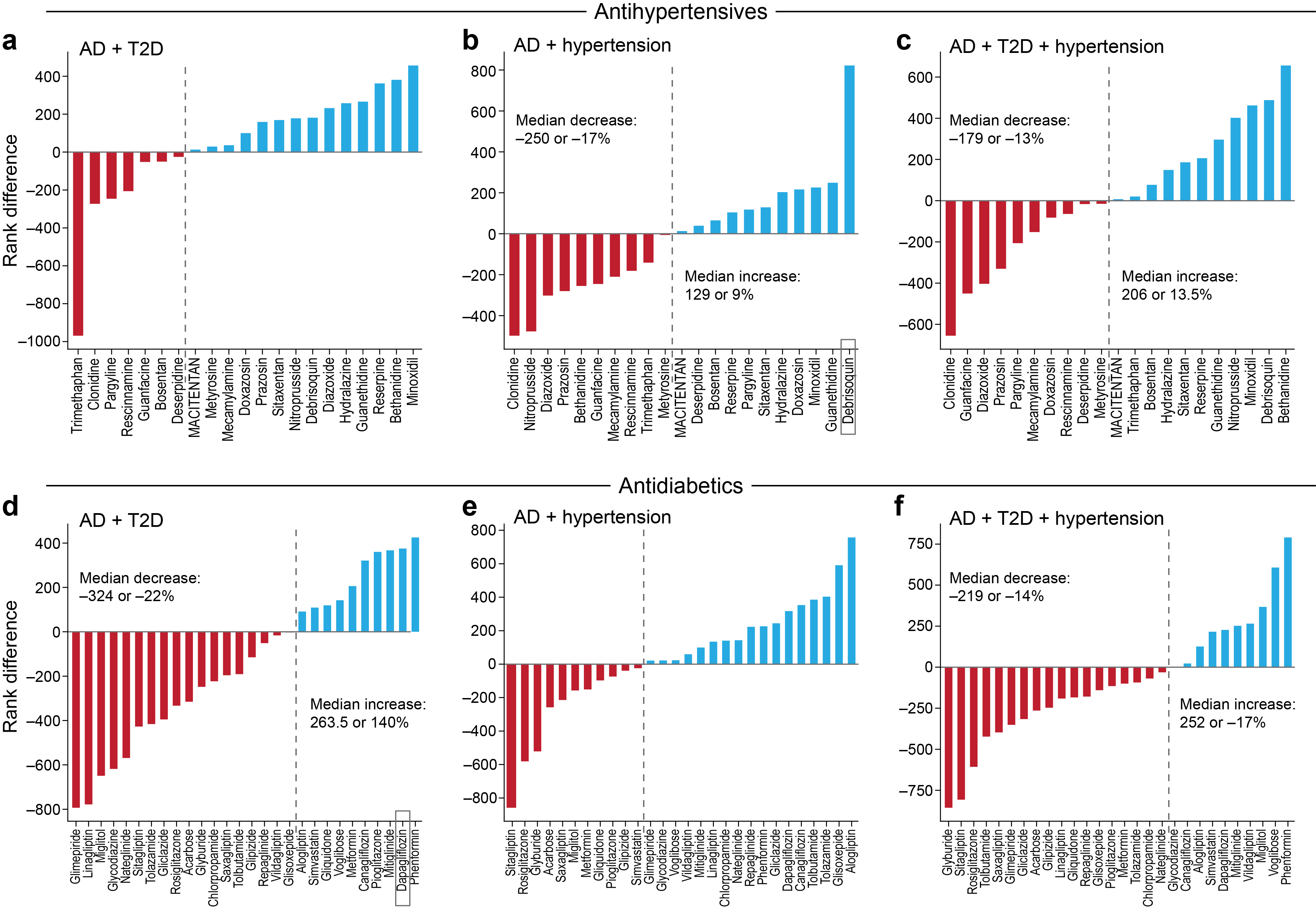}}
\end{figure}

Going forward, we will also leverage large-language models to efficiently query, summarize, and interpret the biomedical literature about each drug that had a significant shift in rank when incorporating the biological pathways from one or two risk factors of the considered disease. 

\section{Additional Details about the Skip-Gram Method}\label{apd:second}

The Skip-Gram model's objective is to devise word representations that effectively predict surrounding words in a sentence or document. Formally stated, given a sequence of $T$ training words $w_1, w_2, ..., w_T$, the model aims to maximize the following average log probability:

\begin{equation}
\frac{1}{T} \sum_{t=1}^{T} \sum_{-k \leq j \leq k, j \neq 0} \log P(w_{t+j} | w_t)
\end{equation}

where $k$ denotes the size of the training context and $T$ denotes the total number of training words. We guided the model to learn embeddings of size $n=64$ and subsequently used cosine similarity as the metric to quantify the similarity between any two biological pathways. 

Our decision to utilize the Skip-Gram algorithm for learning embeddings was driven by our intent to infer semantic contextual relationships among biological pathways, given a specific disease. 

\end{document}